\documentclass[12pt,fleqn]{article}
\usepackage{graphicx}

\usepackage{latexsym}

\usepackage{amsmath}
\usepackage{amsthm}
\usepackage{amssymb}
\usepackage{amsfonts}

\usepackage{color}

\begin{document}

\newcommand{\red}{}

\newcommand{\rf}[1]{(\ref{#1})}
\newcommand{\rff}[2]{(\ref{#1}\ref{#2})}

\newcommand{\ba}{\begin{array}}
\newcommand{\ea}{\end{array}}

\newcommand{\be}{\begin{equation}}
\newcommand{\ee}{\end{equation}}

\newcommand{\const}{{\rm const}}
\newcommand{\ep}{\varepsilon}
\newcommand{\Cl}{{\cal C}}
\newcommand{\rr}{{\vec r}}
\newcommand{\ph}{\varphi}
\newcommand{\R}{{\mathbb R}}  

\newcommand{\e}{{\bf e}}

\newcommand{\m}{\left( \ba{r}}
\newcommand{\ema}{\ea \right)}
\newcommand{\mm}{\left( \ba{cc}}
\newcommand{\miv}{\left( \ba{cccc}}

\newcommand{\scal}[2]{\mbox{$\langle #1 \! \mid #2 \rangle $}}
\newcommand{\ods}{\par \vspace{0.5cm} \par}
\newcommand{\no}{\noindent}
\newcommand{\dis}{\displaystyle }
\newcommand{\mc}{\multicolumn}

\newcommand{\al}{\alpha}
\newcommand{\bb}{\beta}
\newcommand{\ga}{\gamma}
\newcommand{\de}{\delta}

\title{\bf 
Determination of topological properties  of thin samples  \\ 
by the van der Pauw method 
}
\author{
 {\bf Krzysztof R.\ Szyma\'nski\thanks{\footnotesize{e-mail: \tt k.szymanski@uwb.edu.pl}}, Cezary J.\ Walczyk\thanks{\footnotesize{e-mail: \tt walcez@gmail.com }}} 
 \\[0.5ex] and {\bf Jan L.\ Cie\'sli\'nski}\thanks{\footnotesize
 e-mail: \tt j.cieslinski@uwb.edu.pl}, 
\\[0.5ex] {\footnotesize Uniwersytet w Bia{\l}ymstoku,
Wydzia{\l} Fizyki, }
\\ {\footnotesize ul.\ Cio\l kowskiego 1L, 15-245 Bia\l ystok, Poland}
}

\date{}

\maketitle

\begin{abstract}
We solve the problem of determining basic topological properties of flat samples by performing measurements on their outer edge. The global maximum of four probe resistances shows a  characteristic behaviour, which is dependent on the genus (i.e., the number of holes) of the domain. An extension of the van der Pauw method on domains having zero, one, or two holes is presented and discussed. A possibility of measuring topological properties of condensed matter is demonstrated. Experimental results for triply connected domains are presented and explained by continuous symmetry breaking caused by the presence of two holes. The results are consistent with the topological theorem of Hurwitz on the number of automorphisms of Riemann surfaces.
\end{abstract}

\ods



\pagebreak

\section{Introduction}

The van der Pauw technique \cite{P1,P2} is an important and a convenient method of the measurements of two-dimensional homogeneous systems. 
This is {\red a} particular case of four-probe method, revi{\red e}wed recently on the occassion of its 100 anniversary  \cite{Micoli}. The van der Pauw method is especially attractive because of its insensitivity to the sample shape and size, and contact positions. Influence of inhomogeneities on the measurement results were quantitatively characterised in \cite{Koon}.  {\red The p}recision of practical applications can be increased by correction of finite {\red size} of contacts \cite{ Nahlik, Weiss} and direct measurement of contacts resistivity \cite{Gonza}. Methods in which extended contacts are used were presented recently \cite{Auss}. The key assumption of the classical version of van der Pauw method is that the measured domain is simply connected, i.e., there are no holes.

Recently, a generalization was presented for systems with an isolated single hole \cite{SCL}. It was demonstrated that the sheet resistance and the Riemann modulus can be measured \cite{SLC}. Some measurements of a sample with two holes were reported as well, but the research was focused on minimizing the influence of the holes \cite{Sun}. In fact, the main results were presented for contacts placed inside the sample, on the inner boundaries. In the present paper we adopt a different perspective. We are going to detect a presence of holes and, possibly, their number, by performing measurements only on the outer boundary of the sample. 
We consider a conductive thin plate with four different point contacts that are located at points $\al, \bb, \ga$ and $\de$. A current $j_{\al\bb}$ enters the sample at the contact $\al$ and leaves at the contact $\bb$, whereas the potential $V_{\ga\de}$ is measured between contacts $\ga$ and $\de$.  We assume that currents are linear functions of the potentials, i.e., the medium is ohmic. The resistance for contacts $\al, \bb, \ga$ and $\de$ is defined as 
\be  \label{resistance}
R_{\al\bb\ga\de}  = \frac{V_{\ga\de}}{j_{\al\bb}} \   .
\ee

For a simply connected domain (i.e., a domain without holes), the famous van der Pauw equation is satisfied \cite{P1}:
\be  \label{vdP} 
 \exp \left( - \frac{R_{\al\bb\ga\de}}{\lambda} \right) 
 +  \exp \left( - \frac{R_{\bb\ga\de\al}}{\lambda} \right)  = 1
\ee
where $\lambda = \rho/(\pi d)$ and $\rho$ indicates the specific resistivity of the sample with a thickness $d$. The ratio $\rho/d$, known as the sheet resistance, can be determined from Eq.~\ref{vdP} provided that measurements of both resistances ($R_{\al\bb\ga\de}$ and $R_{\bb\ga\de\al}$)  for one arbitrarily chosen location of contacts  are performed. We point out that in the simply connected case $R_{\bb\ga\de\al}$  is a function of $R_{\al\bb\ga\de}$. Therefore, experimental data for a series of measurements should form a line on a graph representing $R_{\al\bb\ga\de}$ versus $R_{\al\bb\ga\de}$, see Fig. 3 in \cite{SCL} and Fig. 1a in \cite{SLC}.  In the multiply connected case the analogical experimental data fill out a two-dimensional region. 
The generalization of the van der Pauw approach for doubly connected domains (i.e., domains with one hole) is presented in \cite{SCL}. For an annulus with radiuses $r_{\rm inn}$ and $r_{\rm out}$, and four contacts located on one edge (outer boundary) at angles $\al, \bb, \ga$ and $\de$, the resistance can be expressed as:
\be  \label{Rabcd} 
R_{\al\bb\ga\de} = \lambda \ln \frac{G(\ga-\al,h)\ G(\de-\bb,h)}{G(\ga-\bb,h)\ G(\de-\al,h)} \ .
\ee
where $h=2\ln \mu$ is a geometric parameter related to the Riemann modulus $\mu = r_{\rm out}/r_{\rm inn}$, whereas $G$ is the function defined by:
\be  \label{G}
  G(\phi,h) = \left|  \sin \frac{\phi}{2}  \right|\ \prod_{n=1}^{\infty} \left( 1 - \frac{\cos\phi}{\cosh h n} \right) \ .
\ee
{\red In the simply connected case $r_{\rm inn} = 0$ which means that $h \rightarrow \infty$. Hence $G(\phi,h) \rightarrow G(\phi,\infty) = |\sin (\phi/2) |$. }
 
The formula  \rf{Rabcd} for resistances is invariant under conformal mapping.  Therefore, it is valid for any domain with a single hole. One of the consequences of the explicit solution \rf{Rabcd}  is the following inequality:
\be  \label{nierownosc} 
 \exp \left( - \frac{R_{\al\bb\ga\de}}{\lambda} \right) 
 +  \exp \left( - \frac{R_{\bb\ga\de\al}}{\lambda} \right)  \leqslant 1 \ .
\ee
In the doubly connected case we have demonstrated that by measuring maximal four-probe resistances it is possible to determine a Riemann modulus \cite{SLC}, whereas by a single measurement of nine four-probe resistances using six contacts it is possible to determine a Riemann modulus and a sheet resistance simultaneously \cite{MST}.  In this paper we show how to determine the presence of more than one hole by performing measurements on the outer edge of the sample.

\section{Equipotential configurations    }

In order to obtain interesting quantitative characterization of topological properties it is convenient to consider only a special class of configurations satisfying the condition $R_{\al\bb\ga\de} = R_{\bb\ga\de\al}$. Note that in the simply connected case this constraint allows a direct computation of  the sheet resistance. Indeed, from Eq.~\rf{vdP} it follows immediately that {\red $R_{\al\bb\ga\de} = \lambda \ln 2$, i.e.,} 
\be   \label{ro}
\rho = \frac{\pi R d}{\ln 2} \ ,
\ee
where $R = R_{\al\bb\ga\de} = R_{\bb\ga\de\al}$. 

A configuration of contacts satisfying the condition $R_{\al\bb\ga\de} = R_{\bb\ga\de\al}$ will be referred to as  equipotential because, due to the reciprocity theorem \cite{Te}, we always have
\be
   R_{\bb\ga\de\al} = R_{\al\bb\ga\de} + R_{\al \ga \de \bb} \ .
\ee
Therefore, the condition $R_{\al\bb\ga\de} = R_{\bb\ga\de\al}$  is equivalent to $R_{\al \ga \de \bb} = 0$ which means that $V_{\de\bb}=0$. The last condition can be easily verified experimentally. Note that choosing arbitrary locations of three contacts on the outer boundary one can always find the location of the fourth contact such that the resulting configuration is equipotential. {\red We point out that Eq.~\rf{ro} obviously implies that in the simply connected case all equipotential configurations yield the same resistance $R_{\al\bb\ga\de}$.}

{\red In the presence of holes (i.e., when the studied sample is ``multiply connected'') we can study dependence of the ``equipotential'' resistance $R_{\al\bb\ga\de}$ with respect to location of contacts. In particular, we consider}  equipotential configurations with maximal resistance. Keeping one contact  at a {\red fixed} position and changing positions of three other contacts (located on the same edge), we were looking for a global maximum of resistance $R_{\al\bb\ga\de}$  under condition $R_{\al\bb\ga\de} = R_{\bb\ga\de\al}$  (or $R_{\al \ga \de \bb} = 0$).  
In the case of an annulus the value of the global maximum of $R_{\al\bb\ga\de}$ under condition $R_{\al\bb\ga\de} = R_{\bb\ga\de\al}$  can be explicitly computed using Eq.\ (5) of \cite{SLC}:  {\red
\be  \label{Gmax}
    R_{\al\bb\ga\de} = 2\lambda \ln   \frac{ G (\pi, h)}{G \left( \frac{\pi}{2}, h \right) } \equiv \lambda \ln 2 + 2\lambda \sum_{n=1}^{\infty} \ln \left( 1 + \frac{1}{\cosh h n} \right) \ .
\ee
 The obtained maximum value of $R_{\al\bb\ga\de}$}  depends only on the Riemann modulus {\red $h$ and $\lambda \equiv \rho/(\pi d)$ (the dependence on $h$ is very strong: $R_{\al\bb\ga\de}$ given by Eq.~\rf{Gmax} tends to infinity for $h\rightarrow 0$). However, t}here is no dependence on the location of the {\red first} contact.
 Ge{\red o}metrically, in the case of annulus, we have a continuum number of such configurations corresponding to squares inscribed into the outer circle. Conformal invariance implies that the same property (independence of the maximum on the location of the first contact) characterizes any doubly connected sample. 

\begin{figure}
\begin{center}
\includegraphics[width=10cm]{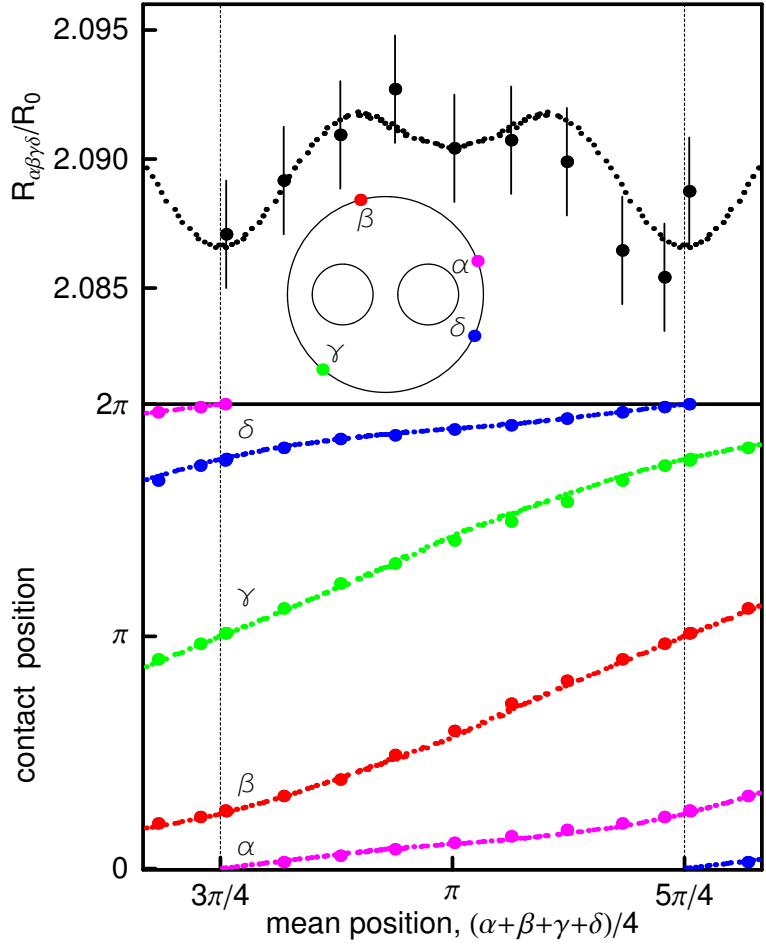}
\end{center}
\caption{  (a) Global maximum of  $R_{\al\bb\ga\de}/R_0$, where $R_0=\lambda \ln 2$  is the global maximum of $R_{\al\bb\ga\de}$  for a sample without a hole. (b) Positions of contacts. Mean position below $3\pi/4$ and above $5\pi/4$  was changed by $\pm \pi/2$  to avoid discontinuity. Large dots are experimental results while small dots  show results of the numerical simulations. Inset: {\red shape of the triply connected sample used in the experiment and} positions of four contacts located at $\al$, $\bb$, $\ga$ and $\de$, related to a global maximum of $R_{\al\bb\ga\de}$ (colors of contacts correspond to colors of dots in (b)). }
\label{Fig mean}
\end{figure}

In the case of two holes the global maximum $R_{\al\bb\ga\de}$ as a function of $\bb, \ga$ and $\de$ under condition $R_{\al\bb\ga\de} = R_{\bb\ga\de\al}$ depends on the position of $\al$, in contrast to the previous cases (disk and annulus). This is observed experimentally (see Fig.~\ref{Fig mean}), where results of measurements are presented conveniently as a function of the mean position of $\al, \bb, \ga$ and $\de$.

\section{Breaking of M\"obius symmetries by holes}

\begin{figure}
\begin{center}
\includegraphics[width=12cm]{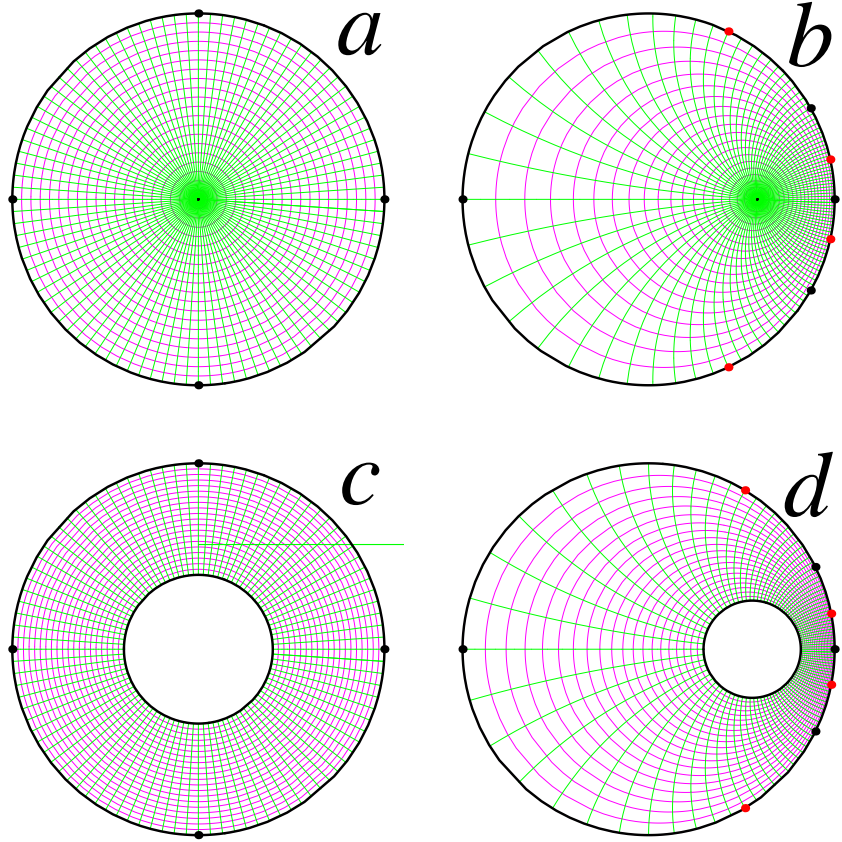}
\end{center}
\caption{(a) example of a disk with four contacts (black points); (b) the disk and contacts shown by black points under M\"obius transformation (3) for parameters $\xi=0.585$  and $\varphi=0$. Red points show the position of contacts under M\"obius transformation for parameters $\xi=0.585 e^{i\pi/4}$ and $\varphi = -\pi/4$. The M\"obius transform{\red ation} is a symmetry of a disk. (c) Example of an annulus with four contacts (black points); (d) the annulus and the position of contacts (black points) under M\"obius transformation (3) for parameters $\xi=0.622$ and $\xi=0$. Red points show the position of contacts under M\"obius transformation for parameters $\xi=0.622 e^{i\pi/4}$ and $\varphi = -\pi/4$.  The M\"obius transform{\red ation} is not a symmetry of an annulus.}
\label{Fig abcd}
\end{figure}

The essence of the van der Pauw method is the invariance of currents and potentials under conformal deformation of a simply connected domain. The simply connected domain is topologically equivalent to a disk. It is known (see, for instance, \cite{Kr}) that any conformal self-map of the unit disk to itself is a M\"obius transformation:
\be   \label{Mob}
    f (z) = \frac{z_0 + z \, {\rm e}^{i \varphi}}{\bar z_0 z \, {\rm e}^{i \varphi} + 1} ,  \quad |z_0|<1 \ , \quad  \varphi \in\R \ ,
\ee
parameterized by one complex ($z_0$) and one real ($\varphi$) parameter (the parameter $\xi$, used in \cite{Kr}, is given by $\xi = z_0 e^{-i\varphi}$). Note that $f(0)=z_0$, i.e., the center of the disk is mapped into $z_0$. It is known that  a M\"obius transformation transforms any circle into a circle (or a straight line), {\red see} Fig.~\ref{Fig abcd}. 
The transformation \rf{Mob} in general is not a symmetry of an annulus mapping it to a disk with an eccentric hole with a center in $z_0$ (see Fig.~\ref{Fig abcd}). The M\"obius symmetries of an annulus reduce to rotations (obtained for $z_0=0$). 

In the case of a disk equipotential configurations are characterized by the cross ratio equal to $\pm 1$ (compare Eq.\ 7 in Ref.~\cite{Ci}). The cross ratio is invariant under M\"obius transformations. Thus any equipotential configuration can be mapped by a M\"obius transform into the square (actually this includes two cases: $+1$ and $-1$, correspond{\red ing} to two different {\red choices of locations} of contacts measuring voltage), compare Fig.~\ref{Fig abcd}ab. It means that for the disk all equipotential configurations have the same four probe resistance \rf{resistance}.  

In the case of the annulus this full symmetry is broken. {\red Equipotential configurations are symmetric with respect to a diameter}.   The four probe resistances  for equipotential configurations are, as a rule, different. {\red The maximum value correspond to the case of contacts located at vertices of a square, see  Fig.~\ref{Fig abcd}c. Applying a M\"obius transformation we obtain the same resistance but for a different sample, see Fig.~\ref{Fig abcd}d.}  The only exception are configuration of the same shape, transformed into each other by rotations of the annulus. The configuration with a maximal resistance \rf{Gmax}, corresponding to the square, is also given up to a rotation.

{\red M\"obius transformations \rf{Mob} preserve the outer disc. We recall that using an appropriate conformal transformation, defined by a holomorphic function  $f(z)$ of more general form, we can transform  a figure of arbitrary shape into the disc. The essence of the van der Pauw approach consists in the fact that four probe resistances are preserved by conformal transformations. }

Any domain with two holes can be conformally mapped on a disk with two concentric {\red slits} \cite{Ah} and then, by using {\red another conformal transformation}, to an annulus with an additional circular hole. The lack of a rotational symmetry of an annulus with an additional circular hole is apparent. One may expect that the global maximum of $R_{\al\bb\ga\de}$  as a function of $\bb, \ga$ and $\de$ under condition $R_{\al\bb\ga\de} = R_{\bb\ga\de\al}$ depends on the position of $\al$, in contrast to two previous cases (disk and annulus). This is observed experimentally (see Fig.~\ref{Fig mean}), where results of measurements are presented as function of the mean position of $\al, \bb, \ga$ and $\de$. Results of our analysis coincide with well-known topological properties of triply connected surfaces with genus $g=2$. According to the Hurwitz theorem for $g>2$, the number of automorphisms is finite \cite{Ac}. A sample with two holes (triply connected, $g=2$) cannot have an infinite number of symmetries (including continuous group of rotational symmetries), in contrast to a sample with a single hole (doubly connected, $g=1$) or a sample without a hole (simply connected, $g=0$). The presence of a second hole breaks the rotational symmetry of the problem.

\section{Experiments and numerical simulations for triply connected samples}

The measurements were performed similarly as those described in \cite{SZ}. The sample was made with the same material as reported in \cite{SCL}; its outer diameter was 290.0(3) mm, and it had two holes of a diameter of 90.0(3) mm with their origins located at distances that were 63.3(3) mm from the sample center {\red (see inset in Fig.~\ref{Fig mean})}. One contact was kept at a {\red fixed} position and by changing positions of three others, located on the same edge, we were looking for a global maximum of resistance $R_{\al\bb\ga\de}$  under condition $R_{\al\bb\ga\de} = R_{\bb\ga\de\al}$. In fact, due to the reciprocity theorem \cite{Te}, this is equivalent to a simpler task of searching for   configurations such that $R_{\al \ga \de \bb} = 0$.  
 A ratio of the specific resistance and the thickness ($\lambda$ in Eq.~\rf{vdP}) was determined in a separate experiment. There is a clear dependence of $R_{\al\bb\ga\de}$ on the position of a specific contact on the circumference (see Fig.~\ref{Fig mean}). For a shape with a single hole or without a hole, the global maximum of $R_{\al\bb\ga\de}$ does not depend on the position of $\al$. Quite large experimental precision, better than 0.2\%, is required for detection of tiny changes in measured resistances. This was achieved using a method described in \cite{SZ}.

Results of measurements were also confirmed by more precise simulations programmed in {\it FreeFem}++ environment \cite{He}. Variational formulation \cite{Ba} was used for numerical solution of two dimensional Laplace equation with the Neumann-type boundary conditions. {\it Gmsh} mesh generator \cite{GR} was used for Delaunay triangulation \cite{GB}. The outer boundary of domain was approximated by polygon having $N$ edges while two holes were approximated by polygons with $N_0$ edges, where $N=720$ and $N_0=223$. The mesh consisted of 87848 triangles. Point contacts of a sink and a source were approximated by one edge of the polygon. Results of simulations are shown in  Fig.~\ref{Fig mean}  by small dots. Full agreement between experiment and numerical simulations was achieved.

{\red According to our earlier papers \cite{SCL,SLC}, an analogical experiment, performed for a doubly connected sample, would yield a straight line parallel to the horizontal axis, namely $R_{\al\bb\ga\de}/R_0 = c_h$, where the constant $c_h$ is given by
\be
    c_h = 1 + \frac{2}{\ln 2} \sum_{n=1}^{\infty} \ln \left( 1 + \frac{1}{\cosh h n} \right) \ .
 \ee 
 The sample with only one hole (of the same size and location as in the 
inset of Fig.~\ref{Fig mean}) has $h \approx 1.850$ and $c_h \approx 1.938$. 
 
In the ``pure'' van der Pauw case (when $h \rightarrow \infty$ and $c_h \rightarrow 1$) we would get the line $R_{\al\bb\ga\de}/R_0 = 1$. Therefore,  experiments of this kind can be used as a test for multiconnectedness of the studied sample. }

The presented method of measurement allows experimental access to topological properties of materials. One may expect that this could be applied in investigations of self-organized two dimensional structures. Possibility of measuring of topological properties may be particularly important in determination of constituents of metamaterials.

\section{Conclusions}

We report new results, which are a step further toward generalization of the potential flows on multiconnected domains that are characterized by the four-probe method. {\red We define equpotential configurations of contacts and then we study  the behaviour of the maximum value of van der Pauw resistances for these configurations.  Performing measurements on the outer boundary, we obtain results  qualitatively different in the simply connected case, doubly connected case and triply connected cases. }

For the first time, we discuss properties of flows {\red studied} by the van der Pauw method on simply, doubly, and triply connected domains while taking into account the symmetry breaking by increased multiconnectivity. 
We have shown the consequences of a M\"obius symmetry of a disk and that the presence of a hole breaks the symmetry. Presence of two holes breaks further {\red the} rotational symmetry. Symmetry breaking has an impact on four-probe resistances, evidenced by experiments and numerical simulations.

The global maximum of the four-probe resistance $R_{\al\bb\ga\de}$  under the condition $R_{\al\bb\ga\de} = R_{\bb\ga\de\al}$ depends on the number of holes of the domain. For a domain without a hole (simply connected domain, genus $g=0$), the maximum of $R_{\al\bb\ga\de}$ equals $\lambda \ln 2$. A single hole breaks a M\"obius symmetry of the system, and the global maximum of $R_{\al\bb\ga\de}$ is given by the formula \rf{Gmax}   (doubly connected domain, $g=1$). The second hole (triply connected domain, genus $g=2$) causes a further symmetry breaking of the domain. No continuous symmetries are present, resulting in the dependence of global four-probe resistance on the position of a selected contact on a sample edge. {\red We point out that} the presented results for the maxima of resistances are invariant under conformal transformations. Therefore, the results {\red obtained and discussed for a disc with circular holes} are valid for any triply connected domain. 

{\red We hope that results of this paper will be useful for designing new experimental approaches in material science. A p}ossibility of measuring topological properties of some structures of condensed matter by electrical methods may have potential application in designing and testing new classes of metamaterials \cite{Nin}.

\ods
\noindent {\bf Acknowledgements.}  The work was supported by funds of Polish Ministry of Science and Higher Education.
The authors would like to thank Piotr Su\l kowski for his valuable discussions and suggestions.


\end{document}